\documentclass[10pt]{article}

\usepackage{amsmath,amssymb}
\usepackage{amsfonts}

\begin{document}
\def\Tr{{\rm Tr }\ }

\def\bC{\mathbb{C}}
\def\CA{{\cal A}}
\def\CB{{\cal B}}
\def\CC{{\cal C}}
\def\CD{{\cal D}}
\def\CE{{\cal E}}
\def\CF{{\cal F}}
\def\CG{{\cal G}}
\def\CH{{\cal H}}
\def\CI{{\cal I}}
\def\CJ{{\cal J}}
\def\CK{{\cal K}}
\def\CL{{\cal L}}
\def\CM{{\cal M}}
\def\CN{{\cal N}}
\def\CO{{\cal O}}
\def\CP{{\cal P}}
\def\CQ{{\cal Q}}
\def\CR{{\cal R}}
\def\CS{{\cal S}}
\def\CT{{\cal T}}
\def\CU{{\cal U}}
\def\CV{{\cal V}}
\def\CW{{\cal W}}
\def\CX{{\cal X}}
\def\CY{{\cal Y}}
\def\CZ{{\cal Z}}

\newcommand{\todo}[1]{{\em \small {#1}}\marginpar{$\Longleftarrow$}}
\newcommand{\labell}[1]{\label{#1}\qquad_{#1}} 
\newcommand{\bbibitem}[1]{\bibitem{#1}\marginpar{#1}}
\newcommand{\llabel}[1]{\label{#1}\marginpar{#1}}

\newcommand{\sphere}[0]{{\rm S}^3}
\newcommand{\su}[0]{{\rm SU(2)}}
\newcommand{\so}[0]{{\rm SO(4)}}
\newcommand{\bK}[0]{{\bf K}}
\newcommand{\bL}[0]{{\bf L}}
\newcommand{\bR}[0]{{\bf R}}
\newcommand{\tK}[0]{\tilde{K}}
\newcommand{\tL}[0]{\bar{L}}
\newcommand{\tR}[0]{\tilde{R}}

\newcommand{\ack}[1]{[{\bf Pfft!: {#1}}]}

\newcommand{\btzm}[0]{BTZ$_{\rm M}$}
\newcommand{\ads}[1]{{\rm AdS}_{#1}}
\newcommand{\ds}[1]{{\rm dS}_{#1}}
\newcommand{\dS}[1]{{\rm dS}_{#1}}
\newcommand{\eds}[1]{{\rm EdS}_{#1}}
\newcommand{\sph}[1]{{\rm S}^{#1}}
\newcommand{\gn}[0]{G_N}
\newcommand{\SL}[0]{{\rm SL}(2,R)}
\newcommand{\cosm}[0]{R}
\newcommand{\hdim}[0]{\bar{h}}
\newcommand{\bw}[0]{\bar{w}}
\newcommand{\bz}[0]{\bar{z}}
\newcommand{\be}{\begin{equation}}
\newcommand{\ee}{\end{equation}}
\newcommand{\bea}{\begin{eqnarray}}
\newcommand{\eea}{\end{eqnarray}}
\newcommand{\pat}{\partial}
\newcommand{\lp}{\lambda_+}
\newcommand{\bx}{ {\bf x}}
\newcommand{\bk}{{\bf k}}
\newcommand{\bb}{{\bf b}}
\newcommand{\BB}{{\bf B}}
\newcommand{\tp}{\tilde{\phi}}
\hyphenation{Min-kow-ski}

\newcommand{\pa}{\partial}
\newcommand{\eref}[1]{(\ref{#1})}

\def\apr{\alpha'}
\def\str{{str}}
\def\lstr{\ell_\str}
\def\gstr{g_\str}
\def\Mstr{M_\str}
\def\lpl{\ell_{pl}}
\def\Mpl{M_{pl}}
\def\varep{\varepsilon}
\def\del{\nabla}
\def\grad{\nabla}
\def\tr{\hbox{tr}}
\def\perp{\bot}
\def\half{\frac{1}{2}}
\def\p{\partial}
\def\perp{\bot}
\def\eps{\epsilon}

\newcommand{\BC}{\mathbb{C}}
\newcommand{\BR}{\mathbb{R}}
\newcommand{\BZ}{\mathbb{Z}}
\newcommand{\bra}[1]{\langle{#1}|}
\newcommand{\ket}[1]{|{#1}\rangle}
\newcommand{\vev}[1]{\langle{#1}\rangle}
\newcommand{\Real}{\mathfrak{Re}}
\newcommand{\Imag}{\mathfrak{Im}}
\newcommand{\talpha}{{\widetilde{\alpha}}}
\newcommand{\Ham}{{\widehat{H}}}
\newcommand{\al}{\alpha}
\newcommand\x{{\bf x}}
\newcommand\y{{\bf y}}

\def\NPB{{\it Nucl. Phys. }{\bf B}}
\def\PL{{\it Phys. Lett. }}
\def\PRL{{\it Phys. Rev. Lett. }}
\def\PRD{{\it Phys. Rev. }{\bf D}}
\def\CQG{{\it Class. Quantum Grav. }}
\def\JMP{{\it J. Math. Phys. }}
\def\SJNP{{\it Sov. J. Nucl. Phys. }}
\def\SPJ{{\it Sov. Phys. J. }}
\def\JETPL{{\it JETP Lett. }}
\def\TMP{{\it Theor. Math. Phys. }}
\def\IJMPA{{\it Int. J. Mod. Phys. }{\bf A}}
\def\MPL{{\it Mod. Phys. Lett. }}
\def\CMP{{\it Commun. Math. Phys. }}
\def\AP{{\it Ann. Phys. }}
\def\PR{{\it Phys. Rep. }}

\renewcommand{\thepage}{\arabic{page}}
\setcounter{page}{1}

\rightline{hep-th/0407051}
\rightline{ILL-(TH)-04-04}
\rightline{VPI-IPPAP-04-06}

\vskip 0.75 cm
\renewcommand{\thefootnote}{\fnsymbol{footnote}}
\centerline{\Large \bf Towards the QCD String:}
\vskip 0.1 cm
\centerline{\Large \bf
$2+1$ dimensional Yang-Mills theory in the planar limit}
\vskip 0.75 cm

\centerline{{\bf Robert G. Leigh${}^{1}$\footnote{rgleigh@uiuc.edu} and
Djordje Minic${}^{2}$\footnote{dminic@vt.edu}
}}
\vskip .5cm
\centerline{${}^1$\it Department of Physics,
University of Illinois at Urbana-Champaign}
\centerline{\it 1110 West Green Street, Urbana, IL 61801-3080, USA}
\vskip .5cm
\centerline{${}^2$\it Institute for Particle Physics and Astrophysics,}
\centerline{\it Department of Physics, Virginia Tech}
\centerline{\it Blacksburg, VA 24061, U.S.A.}
\vskip .5cm

\setcounter{footnote}{0}
\renewcommand{\thefootnote}{\arabic{footnote}}

\begin{abstract}
We study the large $N$ (planar) limit of pure $SU(N)$  2+1 dimensional
Yang-Mills theory ($YM_{2+1}$) using a gauge-invariant matrix
parameterization introduced by Karabali and Nair. This formulation
crucially relies on the properties of local holomorphic gauge invariant collective fields in the
Hamiltonian formulation of $YM_{2+1}$. We show that the spectrum in the
planar limit of this theory can be explicitly determined in the
$N=\infty$, low momentum (large 't Hooft coupling) limit, using the
technology of the Eguchi-Kawai reduction and the existing knowledge
concerning the one-matrix model. The dispersion relation describing the
planar $YM_{2+1}$ spectrum reads as $\omega(\vec{k}) = \sqrt{{\vec{k}}^2
+ m_n^2}$, where $n=1,2,...$ and $m_n = n m_r$, where $m_r$ denotes the
renormalized mass, the bare mass $m$ being determined by the planar 't
Hooft coupling $g_{YM}^2 N$ via $m= \frac{g_{YM}^2 N}{2 \pi}$. The
planar, low momentum limit, also captures the expected short and long
distance physics of $YM_{2+1}$ and gives an interesting new picture of
confinement. The computation of the spectrum is possible due to a
reduction of the $YM_{2+1}$ Hamiltonian for the large 't Hooft coupling
to the {\it singlet} sector of an effective one matrix model. 
The crucial observation is that the correct vacuum
(the large $N$ master field), consistent with the area law and the
existence of a mass gap, is described by an effective quadratic matrix
model, in the large $N$, large 't Hooft coupling limit.

\end{abstract}

\newpage

\section{Introduction}

The study of the large $N$ limit of Yang-Mills theory is one of the grand
problems of theoretical physics. In recent years, a new viewpoint has
emerged concerning the planar limit of gauge theories, mainly motivated
by recent insightful advances in string theory based on the duality
between string and gauge theories \cite{review}. Nevertheless, a precise
formulation (a prerequisite for a solution) of the elusive "QCD string"
is still lacking.

In this letter we take a fresh look at this problem in the setting
provided by $2+1$ dimensional Yang Mills theory ($YM_{2+1}$) - a highly
non-trivial quantum field theory 
\cite{twoplusone}, \cite{knair}. This theory is expected on many
grounds to share the essential features of its $3+1$ dimensional cousin,
such as asymptotic freedom and confinement, yet is distinguished from its $3+1$ dimensional counterpart by the
existence of a dimensionful coupling. We regard this study as a stepping stone towards the $3+1$ dimensional theory.

Interestingly enough, we are able to make a precise statement concerning
the spectrum of this theory in the large $N$, reduced, low momentum
limit. In this limit, we argue, the generic features of the Yang-Mills
vacuum are fully captured by a quadratic large $N$ matrix model. We
perform explicit computations in a well-defined framework utilizing a
momentum expansion of the reduced, planar effective action given in
terms of the local gauge invariant variables which correspond to the
only propagating physical polarization. Our approach utilizes many
recipes from the large $N$ cookbook (the large $N$ reduction, matrix model
technology), and yet is seemingly not {\it directly} related to the
recent advances in the understanding of certain planar gauge theories
from a string theory (gauge theory/gravity duality) point of view. This
of course does not mean that a possible {\it indirect} connection is
non-existent. We note that our approach can be understood as a target
space, Hamiltonian formulation of an effective string field theory
describing the planar $2+1$ dimensional QCD string.

Our work is crucially based on beautiful results derived in the remarkable
work of Karabali and Nair \cite{knair}. They have provided an explicit
gauge invariant reformulation of $YM_{2+1}$ in terms of local
holomorphic variables. On a more practical level, Karabali
and Nair have been able to compute the string tension in their
Hamiltonian approach which is in excellent agreement (up to  3\%) with
the existing lattice data \cite{teper}. That striking result as well as
the computation presented in this letter clearly point out that the
Karabali-Nair approach has some truly remarkable features which can lead
to potentially dramatic results in the arena of $2+1$ dimensional gauge
theories. 

It is reasonable to believe that if a solution of $YM_{2+1}$ is to be found at all, it will be in the planar limit. Consequently, we consider here 
{\em the large $N$ limit} of $YM_{2+1}$ using the Karabali-Nair parameterization, and this is the distinguishing feature of our work.  We consider the spectrum of the
theory and are able to show that there exists a
mass gap set by the 't Hooft coupling. This result extends but is
certainly consistent with the results of \cite{knair}.

The crucial element in the computation of the spectrum is a large $N$
reduction of the $YM_{2+1}$ Hamiltonian, in a well-defined low momentum
limit (large 't Hooft coupling), written in terms of the Karabali-Nair
variables, to the {\it singlet sector of an effective one Hermitian matrix model}.
The singlet sector is selected by the presence of a local holomorphic invariance of the planar
vacuum which arises in the Karabali-Nair formalism. This reduction procedure enables
us to write a self-consistent gap equation for the planar sector of
$YM_{2+1}$. What is most important is that our approach provides a
well-defined momentum expansion of the full effective local gauge
invariant collective field theory of $YM_{2+1}$ in the reduced, planar
limit. In particular, the correct Yang-Mills vacuum ({\em i.e.}, the
large $N$ master field), consistent with the area law and the existence
of a mass gap, is captured by an effective quadratic matrix model, in
the large $N$, large 't Hooft coupling limit. This effective theory of
gauge invariant holomorphic loop variables can be in principle used for
other calculations, such as the determination of various correlation
functions.

The outline of this paper is as follows:
In Section 2 we review the Karabali-Nair variables and then
in Section 3 we
investigate in detail the Karabali-Nair collective field theory Hamiltonian.
The large $N$ limit of this Hamiltonian is studied in Section 4 and the
planar spectrum of the same in Section 5. A few more technical details
related to the analysis of the collective field Hamiltonian are collected
in a separate Appendix at the end of this letter. 

\section{The Karabali-Nair variables}

The Karabali-Nair approach can be summarized as follows \cite{knair}:
consider an $SU(N)$ $YM_{2+1}$ in the Hamiltonian gauge $A_0 =0$. Write
the gauge potentials as $A_i=-it^aA_i^a$, for $i=1,2$, where $t^a$ are
the Hermitian $N\times N$ matrices in the $SU(N)$ Lie algebra $[t^a,
t^b] = if^{abc} t^c$ with the normalization $2Tr(t^at^b) =\delta^{ab}$.
Define complex coordinates $z=x_1-i x_2$ and $\bar{z} =x_1+i x_2$, and
furthermore $2 A^a = A^a_1+iA^a_2$, $2 \bar{A}^a = A^a_1-iA^a_2$.

The Karabali-Nair parameterization is 
\begin{equation} \label{hl}
A= - \partial_z M M^{-1}, \quad \bar{A}= + (M^{-1})^\dagger\partial_{\bar{z}}
M^{\dagger}
\end{equation}
where $M$ is a general element of $SL(N,\BC)$. Note that a (time
independent) gauge transformation $A \to g A g^{-1} - \partial g\
g^{-1}$, $\bar A \to g \bar A g^{-1} - \bar\partial g\ g^{-1}$, where $g
\in SU(N)$ becomes simply $M \to g M$. The variables $M$ correspond to
holomorphic loops; their most important, and perhaps unexpected,
property is locality! The corresponding {\it local gauge invariant} variables are given in terms
of ``closed loops" $H \equiv M^{\dagger} M$. Note that the standard Wilson loop operator
may be written
\begin{equation}\Phi(C)=Tr P exp \{ -i\oint_C dz\ \partial_z H H^{-1}\}\end{equation}
and thus is closely related to the local $H$ variables. 

Now one might wonder whether the parameterization (\ref{hl}) is well-defined.
In fact, the definition of $M$ implies a {\it holomorphic invariance} 
\begin{eqnarray}
M(z,\bar z) &\to& M(z,\bar z)h^\dagger (\bar{z})\\
M^\dagger(z,\bar z) &\to& h({z})M^\dagger(z,\bar z)
\end{eqnarray}
where $h(z)$ is an arbitrary unimodular complex matrix whose matrix
elements are independent of $\bar z$. Under the holomorphic
transformation, the gauge invariant variable $H$ transforms
homogeneously
\begin{equation}
\label{holo}
H(z,\bar z) \to h(z) H(z,\bar z) h^\dagger(\bar{z})
\end{equation}
This is distinct from the original gauge transformation, since it acts
as right multiplication rather than left and is holomorphic. One way to
understand its appearance is that the parameterization (\ref{hl}) can be
formally inverted in the form
\begin{equation}
M(x,\bar x)=\left( 1-\int d^2z\ G(x,z)A_z(z,\bar z)+\ldots\right) \bar V(\bar x)
\end{equation}
where $G$ is the Green's function, $\partial_z G(z,x)=\delta^{(2)}(z-x)$
and $\bar V$ is an arbitrary matrix with only anti-holomorphic
dependence. The theory written in terms of the gauge invariant $H$
fields will have its own local (holomorphic) invariance. The gauge
fields, and the Wilson loop variables, know nothing about this extra
invariance. We will deal with this, as in \cite{knair}, by requiring
that the wave functions (or equivalently, physical states) be
holomorphically invariant. The insistence of the holomorphic invariance
of the vacuum is of crucial importance for our main argument in what
follows.

One of the most remarkable properties of this parameterization is that
the Jacobian relating the measures on the space of connections $C$ and
on the space of gauge invariant variables $H$ can be explicitly
computed
\begin{equation}
d \mu [C] = \sigma d \mu [H] e^{2c_A S_{WZW}[H]}
\end{equation}
where $c_A$ is the quadratic Casimir in the adjoint representation of
$SU(N)$ ($c_A = N$) and
\begin{equation}
S_{WZW}(H)= - \frac{1}{2\pi}\int d^2z\ Tr H^{-1}\partial H H^{-1}\bar\partial H+
\frac{i}{12\pi} \int d^3x\
\epsilon^{\mu\nu\lambda} Tr H^{-1}\partial_\mu HH^{-1}\partial_\nu
HH^{-1}\partial_\lambda H
\end{equation}
is the level $-c_A$ hermitian Wess-Zumino-Witten action, which is both
gauge and holomorphic invariant. $\sigma$ is a constant determinant
factor.
Thus the inner product may be written as an overlap integral of gauge
invariant wave functionals with non-trivial measure
\begin{equation}
\langle1|2\rangle= \int d \mu [H] e^{2c_A S_{WZW}(H)} \Psi^*_1 \Psi_2
\end{equation}

The standard $YM_{2+1}$ Hamiltonian
\begin{equation}
\int Tr \left(g_{YM}^2 {E_i}^2 + \frac{1}{g_{YM}^2} {B}^2\right)
\end{equation}
can be also explicitly rewritten in terms of gauge invariant variables.
The collective field form \cite{collective} of this Hamiltonian (which
we will refer to as the Karabali-Nair Hamiltonian) can be easily
appreciated from its explicit form in terms of the natural
``current''-like gauge invariant variables\footnote{The $J$ variables
transform as connections under the holomorphic transformation.}
$J=\frac{c_A}{\pi}\partial_z H H^{-1}$,
\begin{equation}
{\cal H}_{KN}[J]=m \left(\int_x J^a(x) \frac{\delta}{\delta J^a(x)} + \int_{x,y}\Omega_{ab}(x,y)
\frac{\delta}{\delta J^a(x)} \frac{\delta}{\delta J^b(y)}\right) +
\frac{\pi}{m c_A} \int_x \bar{\partial} J^a \bar{\partial} J^ a
\end{equation}
where 
\begin{equation}
m = \frac{g_{YM}^2 c_A }{2 \pi}, \quad \Omega_{ab}(x,y) = \frac{c_A}{\pi^2}
\frac{\delta_{ab}}{(x-y)^2} - \frac{i}{\pi} \frac{f_{abc}J^c(x)}{(x-y)}.
\end{equation}
The derivation of this Hamiltonian involves carefully regulating certain
divergent expressions in a gauge invariant manner. We note that the
scale $m$ is essentially the 't Hooft coupling.

At this point we remind the reader about the difference between
collective field theory and effective field theory. Collective field
theory \cite{collective} is simply based on a choice of
collective variables, appropriate to the physics in question. The
technical difficulty usually lies in the explicit change of variables,
which generically renders the collective field theory horribly
non-local. (This is, for example, the case with the collective field
theory of canonical Wilson loop variables.) The crucial requirement that
the large $N$ collective field theory has to meet is the factorization
of vacuum correlators. The factorization in turn implies, by the
resolution of the identity, that the only state that controls the
physics at large $N$ is the vacuum. Notice that because of factorization
at large $N$, one essentially has to be concerned with the appropriate
classical phase space of gauge invariant observables and their
canonically conjugate partners and correspondingly the classical
Hamiltonian. The expectation values (evaluated using appropriate
semiclassical coherent states) of the quantum Hamiltonian, lead to the
required classical Hamiltonian, which in turn is nothing else but the
collective field Hamiltonian \cite{collective}.

The truly amazing feature of the Karabali-Nair holomorphic loop
variables is that they are local and that the
corresponding Jacobian can be explicitly computed! This Jacobian,
determined in terms of the Wess-Zumino-Witten action, enjoys certain
analyticity properties which render it unique and independent
of regularization ambiguities.

The passage to the collective field Hamiltonian may be
thought of as a starting point: having performed the change of
variables, we may then analyze the theory using effective field theory
techniques. In particular, one expects in general that there will be
renormalizations; because of the local nature of the variables, one may
expect that a suitable perturbative analysis can be found which sensibly
deals with such matters. Indeed, we will describe such a formalism here
and explain how the dynamics of the mass gap and confinement arises.

Finally, note that the inner product can be put into a canonical form
\begin{equation}
\langle1|2\rangle= \int d \mu [H] \Phi^*_1 \Phi_2
\end{equation}
provided we perform a redefinition of wavefunctionals $\Phi = e^{{c_A}
S_{WZW}(H)} \Psi$. In so doing, there will be a corresponding adjustment
of the collective Hamiltonian, containing new terms. We will display
this explicitly in the following sections, but note here that the most
important effect is to add a term of the form\footnote{This comes from a
piece of $S_{WZW}$.} $m^2 Tr(\partial H \bar{\partial} H^{-1})$, which
will later be understood to correspond to the appearance of the mass gap
in the large $N$ limit. Of course, near $g_{YM}\to 0$, we would take the
perturbative vacuum and this term is of no particular relevance --- the
description of the theory in terms of $A$ and $\bar A$ is adequate.
However, this term has arisen from the Jacobian of the path integral
measure and although the change of variables in the measure is
essentially an operation on the {\em classical} configuration space, we
claim that one is lead to address the physical {\em non-perturbative}
vacuum. It is with respect to this vacuum that the mass gap appears.

One notes that this collective field formalism is true for any rank of
the gauge group, and in particular agrees with the large $N$ 't Hooft
counting. We will obtain additional insight into the dynamics of
confinement however by examining the theory in the large $N$ limit.
First, let us continue reviewing the results of Refs. \cite{knair}.

\subsection{The vacuum wave functional, area law and string tension}

One of the major results of the Karabali-Nair collective field theory
approach is the analytic deduction of the area law and an explicit
computation of the string tension \cite{knair}.

The computation of the string tension is achieved by an approximate
formula for the vacuum wave functional $\Psi$. First, one notices that
$\Psi=1$ is annihilated by the kinetic term and is normalizable given
the non-trivial measure (due to the normalization of the WZW path
integral).

One may find a vacuum wave functional annihilated by the total collective
Hamiltonian (${\cal H} \Psi \equiv (T+V)\Psi =0$) by expanding this
equation in powers of (roughly) $B/m^2$. To leading order, one finds
\begin{equation}\label{eq:KNwavefn}
\Psi \simeq \exp\left[- \frac{1}{2g_{YM}^2}
\int B(x) \left(\frac{1}{m + \sqrt{m^2 - \nabla^2}}\right) B(y)\right]
\end{equation}
Note that this wavefunctional apparently interpolates between the low and
high momenta regions. At high momenta, this wavefunction correctly has a
form corresponding to free gluons $\Psi\sim e^{-\frac{1}{2g_{YM}^2}\int
B^2/k}$, appropriate to the conformally invariant two-point function of
gluons, $\langle AA\rangle\sim g^2_{YM}/|k|$. In the low momentum
region, the momentum factor is cut-off, and $B^2/k\to B^2/m$. Although
higher order corrections are expected to be non-local, this can make
sense self-consistently, if the theory may be re-organized into an
expansion in inverse powers of $m$.

As explained in \cite{knair}, the low momentum limit  
\begin{equation}
\Psi = \exp\left(- \frac{1}{2g_{YM}^2m}\int \Tr B^2\right)
\end{equation}
provides a probability measure $\Psi^* \Psi$ 
equivalent to the partition function
of the Euclidean two-dimensional Yang-Mills theory
with an effective Yang-Mills coupling $g_{2D}^2 \equiv m g_{YM}^2$.
Using the results from \cite{2dym}, Karabali, Kim and Nair deduced the
area law for the expectation value of the Wilson loop operator
\begin{equation}
\langle\Phi\rangle \sim \exp( - \sigma A)
\end{equation}
with the string tension following from the results of \cite{2dym}
\begin{equation}
\sigma = g_{YM}^4 \frac{N^2 - 1}{8\pi}
\end{equation}
This formula agrees beautifully with extensive lattice simulations
\cite{teper}, and is certainly consistent with the appearance of a mass
gap. Notice that this result is once again in full agreement with the
large $N$ 't Hooft expansion.

\section{The collective field Hamiltonian}

We are interested in studying the planar limit of the Karabali-Nair
approach to $YM_{2+1}$. The large $N$ limit is expected to be controlled
by a constant $\infty \times \infty$ matrix configuration called the
``master field'' \cite{witten}. Such a configuration should capture
correctly both the short and long distance properties, that is, both
asymptotic freedom as well as confinement. As already mentioned above,
knowing the master field configuration is, in a very precise sense,
equivalent to knowing the correct vacuum at large $N$, which is the most
remarkable result of the Karabali-Nair approach as seen from the
preceding section. Small perturbations around this configuration should
lead to the spectrum of glueballs. In the planar limit this spectrum is
expected to be equidistant, consisting of an infinite number of
non-interacting massive colorless particles.

Can we compute the planar spectrum of $YM_{2+1}$ using the Karabali-Nair
scheme? The claim of this letter is that in the low momentum, or
equivalently large 't Hooft coupling limit, the planar spectrum of
$YM_{2+1}$ is explicitly computable. Because the knowledge of the master
field is equivalent to the knowledge of the true vacuum in the large N
limit, the crucial property to be used is of the holomorphic invariance
of the gauge invariant variables $H \to V H \bar{V}$. This fact, in
combination with the known properties of the spectrum of singlet states
in the one matrix model is what makes the computation of the planar
spectrum possible.

The crucial observation we make here is that the above Karabli-Nair
vacuum wave functional consistent with the area law (and as we will see,
with the existence of a mass gap), is captured by an effective quadratic
matrix model, in the large $N$, large 't Hooft coupling limit. The usual
power counting arguments (based on the power expansion in terms of the
large $N$ 't Hooft coupling) can be applied to the Karabali-Nair vacuum
wave functional by assuming a WKB ansatz, $\Psi = \exp(\Gamma)$, which
is consistent with factorization at large N. By assuming a local
expansion of $\Gamma$ in terms of gauge invariant observables, such as
$J$ currents, (involving possible non-local, $J$-independent kernels),
the leading term in the large $N$ 't Hooft coupling is given by a
quadratic expression in terms of currents! We will show that this
Karabali-Nair vacuum wave functional is reproduced by an effective
quadratic matrix model involving renormalized couplings (such as the
mass $m$).

In order to get to this result our first aim is to better understand the
structure of the Karabali-Nair collective field Hamiltonian. That is the
subject of the present section. In the following section we will study
the collective field Hamiltonian in the planar limit.

The Karabali-Nair collective field Hamiltonian can be thought
of as a string field theory Hamiltonian for a pure QCD string in $2+1$
dimensions. Indeed, the Karabali-Nair variables represent local gauge invariant
variables and act operatorially on a true, non-perturbative
Fock space of $2+1$ dimensional Yang-Mills theory. Thus this description
is intrinsically second-quantized and gauge invariant, and as such
does qualify as a QCD string field theory. Note that this
second quantized theory does act as an interacting theory in the usual
space, in other words this QCD string field theory is not
formulated on a loop space, precisely because the string field in this
approach (identified with the Karabali-Nair variables) is local.
As we will see in the next section, the Hamiltonian of this effective QCD
string field theory is generically non-local.
A first quantized worldsheet theory remains for the moment elusive;
presumably such a theory is interacting, and
moreover, can be deduced from the target space second quantized 
theory studied in this letter. Nevertheless, a purely first quantized description should not be considered satisfactory
as it would yield just the spectrum and compute S-matrix-type observables.
The second quantized formulation in principle contains full knowledge of
the non-perturbative Fock space.  In this letter we concentrate on the
form of the non-perturbative planar vacuum and the spectrum of gauge invariant
excitations around it.

\subsection{Hamiltonian}

The classical mechanics of the $H$ variables is somewhat complicated by
the constraint $\det H=1$. The Hamiltonian is more easily found using
traceless variables -- for example, it is convenient to use the currents
$J=\frac{c_A}{\pi} \partial H H^{-1}$. The Karabali-Nair
Hamiltonian\cite{knair} is
\begin{equation}\label{eq:HamKN}
{\cal H}_{KN}[J]=m\int J^a\frac{\delta}{\delta J^a}+m\int_{x,y} 
\Omega^{ab}(x,y) \frac{\delta}{\delta J^a(x)}\frac{\delta}{\delta J^b(y)} +
\frac{\pi}{m c_A} \int \bar{\partial} J^a \bar{\partial} J^a
\end{equation}
where 
\begin{equation}
\Omega^{ab}(x,y)=\frac{c_A\delta^{ab}}{\pi^2 (x-y)^2}-i\frac{f_{abc}J^c(y)}{\pi(x-y)}
\end{equation}
This is derived within a consistent gauge-invariant regularization
scheme. The last term in (\ref{eq:HamKN}) is the potential term, and
follows from the precise relation $\bar\partial J=\frac{c_A}{2\pi i}
M^\dagger BM^{-\dagger}$.

Alternatively, suppose we consider expanding around the constant
solution as
\begin{equation}H=e^{\varphi},\end{equation} 
where $\varphi=\varphi^a t^a$ is Hermitian and traceless. The perturbative vacuum is described by
$H=1$, and we are to expand the theory in powers of $\varphi$. Given the
parameterization of $H$, one finds
\begin{equation}
H^{-1}\partial H=\partial\varphi+\frac12 \left[\partial\varphi,\varphi\right]+
\frac16\left[\left[\partial\varphi,\varphi\right] \varphi\right]+\ldots
\end{equation}
Using the (adjoint) notation\footnote{We distinguish the two
representations by explicit indices in the following formulae when
necessary. Note that here $\varphi^{ab}$ is real skew-symmetric.}
$\varphi^{ab}\equiv \varphi^c f^{abc}$ (or equivalently,
$f^{abc}\varphi^{bc}=c_A\varphi^a$), we find 
\begin{equation}
H^{-1}\partial H=t^a\partial\varphi^b e_{ba}{[\varphi]}
\end{equation}
where\footnote{There is an explicit resummation
$e_{[\varphi]}=i\varphi^{-1}(e^{-i\varphi}-1),$ in the adjoint
notation.} $e_{ba}{[\varphi]}=\delta_{ba}-\frac{i}{2}
\varphi_{ba}-\frac16(\varphi^2)_{ba}\ldots$ is a functional of
$\varphi$. Note that in this notation, we have
a generalized non-Abelian bosonization formula
\begin{equation}\label{eq:Jphi}
J^a [\varphi]= \frac{ic_A}{\pi} e_{ab}[\varphi] \partial\varphi^b.
\end{equation}
Moreover one can also establish the following useful formula (see the Appendix):
\begin{equation}\frac{\delta}{\delta J^a(x)}=\frac{i\pi}{c_A}\int_y {{D'}^{-1}_{ac}}_{(x,y)}\frac{\delta}{\delta\varphi^c(y)}\end{equation}
where
${D'}_{ab}\simeq\delta_{ab}\partial-\frac{i}{2}f_{abc}\varphi^c\partial+if_{abc}\partial\varphi^c+\ldots$.

Given this dictionary between $J$ and $\varphi$ variables,
the collective field theory Hamiltonian may also be expressed in terms of the $\varphi$ variables. It has the form
\begin{eqnarray}\label{eq:HamKNphi}
{\cal H}_{KN}[\varphi]&=&\int_{x}P^a[\varphi](x)\frac{\delta}{\delta\varphi^a(x)}
+\int_{x,y} Q^{ab}[\varphi](x,y)\frac{\delta}{\delta\varphi^a(x)}
\frac{\delta}{\delta\varphi^b(y)}+
\frac{\pi}{m c_A} \int_x \bar{\partial} J^a [\varphi] \bar{\partial} J^a [\varphi]
\end{eqnarray}
The formulae for the functionals $P^a[\varphi]$ and $Q^{ab}[\varphi]$
can be found in the Appendix. Formally, we find point-split versions of:
\begin{eqnarray} 
P^a[\varphi](x)&=&-\frac{g_{YM}^2}{2}\int_{z} (e^{-1}(x))_{ce}\bar G_{(x,z)} 
H(z)_{ed}G_{(z,x)}(e^{-1}(x))_{da,c}\\
Q^{ab}[\varphi](x,y)&=&-\frac{g_{YM}^2}{2}\int_{z} (e^{-1}(y))_{be}
\bar G_{(y,z)}H(z)_{ed}G_{(z,x)}(e^{-1}(x))_{da}
\end{eqnarray}
As in the Appendix, $H_{ab}$ is the adjoint representation
$(e^{-i\varphi})_{ab}$. As we will see, the change of variables
from $J$ to $\varphi$ does introduce extra technical problems in the regularizaton of various expressions, but these
issues do not effect the final physical result.
In particular, the mass gap is not
an artefact of regularization, precisely because the collective
Hamiltonian was derived using a consistent gauge-invariant 
regularization scheme.

\subsection{WZW}
 
Next, let us look at the WZW action. We have
\begin{equation}
S_{WZW}=-\frac{1}{2\pi}\int_M 
Tr\ H^{-1}\partial H. H^{-1}\bar\partial H
+\frac{i}{12\pi}\int_P \epsilon^{\mu\nu\lambda} 
Tr\ H^{-1}\partial_\mu H.H^{-1}\partial_\nu H.H^{-1}\partial_\lambda H
\end{equation}
where $\partial P=M$ and $H$ is Hermitian. 
Given the parameterization of $H$, we can then evaluate
\begin{eqnarray}
Tr\ H^{-1}\partial H.H^{-1}\bar\partial H&=&
\frac12\partial\varphi^a\bar\partial\varphi^b e_{ac}[\varphi]e_{bc}[\varphi]\\
&=&\frac12\partial\varphi^a\bar\partial\varphi^b g_{ba}[\varphi]
\end{eqnarray}
where  $g[\varphi]=e_{[\varphi]} e_{[-\varphi]}=I-\frac{1}{12}\varphi^2+\ldots$ 

The WZW term may be written 
\begin{equation}
-\frac{i}{4\pi}\int d^2zds\ h_{abc}[\varphi] 
\partial\varphi^a\bar\partial\varphi^b\partial_s\varphi^c
\end{equation}
where $h_{abc}=f_{def}e_{ad}e_{be}e_{cf}$. This term is of course a total derivative, and gives
\begin{equation}
\frac{i}{4\pi}\int d^2z\ \partial\varphi^a\bar\partial\varphi^bb_{ba}[\varphi] 
\end{equation}
where $h_{abc}=b_{ab,c}-b_{cb,a}-b_{ac,b}$. ($b=\frac{1}{3}\varphi+\ldots$ is antisymmetric).

Thus we arrive at the familiar result that the full WZW action can be
written as a sigma model action
\begin{equation}
S_{WZW}= -\frac{1}{4\pi}\int \partial\varphi^a\bar\partial\varphi^b G^{WZW}_{ab}(\varphi)
\end{equation}
where
\begin{equation}
G^{WZW}_{ba}(\varphi)=\left( g+ib\right)_{ba}(\varphi)
\end{equation}
For small 
$\varphi$ fields, which correspond to $H$ of order one, one can
use the standard background field method. We obtain
\begin{equation}
G^{WZW}_{ba}(\varphi)=\left( 1+\frac{i}{3}\varphi-\frac{1}{12}\varphi^2+\ldots\right)_{ba}
\end{equation}
The calculations presented below can be obtained by following this
formalism.

Now, we may proceed to do the redefinition of the wavefunctionals. The
effect of this redefinition is to give a new collective Hamiltonian,
which is a similarity transform of ${\cal H}_{KN}$
\begin{equation} {\cal
H}'=e^{c_AS_{WZW}}{\cal H} e^{-c_AS_{WZW}} \equiv {\cal H}_2 + {\cal
H}_1 + {\cal H}_0 
\end{equation} 
where
\begin{equation}
{\cal H}_2 = \int_{x,y} Q^{ab}[\varphi](x,y)\frac{\delta}{\delta\varphi^a(x)}
\frac{\delta}{\delta\varphi^b(y)}
\end{equation}
and
\begin{equation}
{\cal H}_1 = \int_x\left[ P^a[\varphi](x)-2c_A\int_y Q^{ab}[\varphi](x,y)
\frac{\delta S_{WZW}}{\delta\varphi^b(y)}\right]\frac{\delta}{\delta\varphi^a(x)}
\end{equation}
and, finally
\begin{eqnarray}
{\cal H}_0 = -
\frac{c_A}{m \pi} \int_x \bar{\partial} (e_{ab}\partial\varphi^b) 
\bar{\partial} (e_{ac}\partial\varphi^c)-c_A\int_xP^a[\varphi](x)
\frac{\delta S_{WZW}}{\delta\varphi^a(x)}\\-c_A\int_{x,y}Q^{ab}[\varphi](x,y)
\left[\frac{\delta^2 S_{WZW}}{\delta\varphi^a(x)\delta\varphi^b(y)}
-c_A\frac{\delta S_{WZW}}{\delta\varphi^a(x)}\frac{\delta S_{WZW}}{\delta\varphi^b(y)}\right]
\end{eqnarray}
These are formal manipulations and one needs to take care to regulate
the expressions in a gauge and holomorphic invariant manner. In fact, in
general this expression will be {\it non-local}; however, as we will see
later, the real utility of the $\varphi$ variables is that the Jacobian
discussed above essentially generates a mass gap. 
The mass gap is set by the 't Hooft coupling
itself, and so there should be a self-consistent low-momentum expansion,
in powers of $k/m$.

Let us work in an expansion in powers of $\varphi$. Recall that
$e_{ba}=\delta_{ba}-\frac{i}{2}
\varphi_{ba}-\frac16(\varphi^2)_{ba}\ldots$ and
$G_{ab}=(g+ib)_{ab}=\delta_{ab}+\frac{i}{3}\varphi_{ab}+\ldots$. Note
that we have
\begin{eqnarray}
\frac{\delta S_{WZW}}{\delta\varphi^a(x)} &=&
\frac{1}{2\pi}\left[\bar\partial\partial\varphi^b g_{ab}[\varphi]
+\frac{1}{2}\partial\varphi^c\bar\partial\varphi^d 
(G_{ad,c}[\varphi]+G_{ca,d}[\varphi]-G_{cd,a}[\varphi])\right]\nonumber\\
&=&\frac{1}{2\pi}\left[ \bar\partial\partial\varphi^a
+\frac{i}{2} f_{abc}\partial\varphi^c\bar\partial\varphi^b+\ldots
\right]
\end{eqnarray}
Using this expression, we may derive, as an expansion in $\varphi$,
\begin{eqnarray}\label{eq:finalham}
{\cal H}'= - \frac{g_{YM}^2}{2}\int_x \pi^a(x)\int_yC(x,y)\pi^a(y)
+\frac{m_r^2}{2g_{YM}^2}\int\partial\varphi^a\bar\partial\varphi^a
+ \frac{2}{g_{YM}^2} \int \bar{\partial}
\varphi^a
(-\partial \bar{\partial}) \bar{\partial} \varphi^a+\ldots
\end{eqnarray}
where
$C(x,y)$ is in general a non-local kernel whose inverse is, to leading order in $k$, $|k|^2$.

This collective Hamiltonian ${\cal H}'$ is also in general non-local. The terms given in eq. (\ref{eq:finalham}) represent the part of the Hamiltonian which controls the ground-state in the planar limit; we note however that
the parameter $m$ should be understood as a renormalized
coupling $m_r$. 
The quadratic terms determine a dispersion
relation of the form
\begin{equation} 
\Delta(k)=|k|^2\left( E^2-|k|^2-m_r^2\right).
\end{equation}
This indicates that the field $\varphi$ is not canonically normalized. However, since $\varphi$ has no physical
zero-mode\footnote{Constant $\varphi$ is equivalent, via a constant
holomorphic transformation, to $\varphi=0$.}, the transformation to
canonically normalized excitations is non-singular, albeit non-local. We
will explore this further in the next section.

\section{$N=\infty$ reduction}

All of the above analysis is valid at any $N$. Let us now consider the
theory in the planar limit. For self-consistency of our presentation we
briefly review the Eguchi-Kawai reduction \cite{reduction}, which is
strictly valid in the $N=\infty$ limit. We discuss the reduction in the
continuum and concentrate on the matrix scalar field theory, obviously
relevant for our discussion.

Consider the following general local action for a scalar matrix field
$\varphi$
\begin{equation}
S(\varphi) = N\int \Tr[\frac{1}{2} (\partial_a \varphi)^2 + V(\varphi(x))]
\end{equation}
The $N=\infty$ reduction is captured by the following recipe:
\begin{quote}
1) Replace the position dependent scalar matrix field $\varphi(x)$ by
\begin{equation}
\varphi(x) \to e^{i P_a x^a} \varphi_R e^{-i P_a x^a}
\end{equation}
and
\begin{equation}
\pi(x) \to e^{i P_a x^a} \pi_R e^{-i P_a x^a}
\end{equation}
where $P_a$ is a diagonal Hermitian matrix $P^a = diag(p_1^a,
p_2^a...p_N^a)$ and

2) replace the derivative operation $\partial_a\varphi(x)$ by
\begin{equation}
\partial_a \varphi(x) \to i [P_a, \varphi_R]
\end{equation}
and

3) finally, replace the continuum action per unit space-time volume by
\begin{equation}
S_R = Tr[ -\frac{1}{2} [P_a, \varphi_R]^2 + V_R(\varphi_R)]
\end{equation}
\end{quote}
Then the correlation functions of the continuum theory (in the
$N= \infty$ limit) are
given by the correlation functions of the reduced theory,
after integrations over the eigenvalues of the momenta $p_i$
\begin{equation}
\langle F[\varphi(x)]\rangle \to \int \prod_{i=1}^N d^D p_i
\langle F[e^{i P_a x^a} \varphi_R e^{-i P_a x^a}]_R\rangle
\end{equation}
Following this prescription we can easily write a reduced form of
the local part of the collective field Hamiltonian in terms of
$\varphi$ variables.
This is sufficient, because this part of the collective field
Hamiltonian controls the planar vacuum at large 'tHooft coupling,
as we will see in what follows.

As we would like to go to large $N$, we consider the continuum
Eguchi-Kawai reduction of the approximate local expression of the
Hamiltonian (\ref{eq:finalham}).
The low momentum limit
\begin{equation}
\frac{p_i^2}{m^2} \ll 1
\end{equation}
{\it defines} the large 't Hooft limit, given the fact that the gauge
coupling in $2+1$ dimensions is dimensionful. As we noted above, in the
large $N$, large 't Hooft coupling limit, the vacuum wave functional is
given by an exponent of a quadratic functional in the current variables.
 We will show that the large $N$ large 't Hooft coupling limit is
self-consistent in the sense that it leads to the correct vacuum
implying a gap in the spectrum.

To proceed, we introduce the following change of variables in momentum
space (a non-local change of variables in real space) that we alluded to
in the previous section, in order to get a canonically
normalized\footnote{Note that we normalize to 1 rather than $N$ here for
simplicity.} kinetic term
\begin{equation}
\label{nonlocal}
\phi^a(\vec{k})= \sqrt{\frac{k \bar{k}}{g_{YM}^2}}\ \varphi^a (\vec{k}).
\end{equation}
We see that in the low momentum limit of
the reduced Hamiltonian we simply get
\begin{equation}\label{eq:matrixfieldham}
\frac{1}{2} \int Tr \left( \pi^2 +
m_r^2 \phi^2 -
\phi [P_a, [P^a, \phi]]+...\right)
\end{equation}
where the canonical momentum $\pi = - i \frac{\delta}{\delta \phi}$.

As it stands this Hamiltonian apparently describes $N^2-1$ massive
degrees of freedom with a mass proportional to the square of the gauge
coupling, as noted originally in \cite{knair}. Obviously this does not
seem to give the  confining spectrum we expect of $YM_{2+1}$! As a
matter of fact, in the limit of the zero Yang-Mills coupling the
familiar perturbative spectrum of gluons is readily recovered, as we
discussed in Section 2. This result is natural as we have expanded $H(x)
= \exp(\varphi(x))$ around the ``perturbative vacuum'' $H=1$ to
quadratic order in $\varphi$. The new interesting feature in this
discussion is the presence of the gauge invariant mass term, whose
origin was the Jacobian of the transformation to these variables.

Nevertheless we claim that one can gain important insight into the
nature of the planar vacuum provided we remember that the master field,
in terms of gauge invariant $H = \exp(\varphi)$ variables, is supposed
to be a {\it constant} $\infty \times \infty$ matrix that transforms
homogeneously under the residual {\it constant} transformations $H \to h
H h^{\dagger}$. In particular, {\em the vacuum is preserved by constant
unitary transformations}. Therefore in the planar limit the matrix
$\phi$ will also be an $\infty \times \infty$ matrix that transforms as
$\phi \to h \phi h^{\dagger}$. We must require invariance under constant
unitary transformations on the zero-mode of this field.\footnote{This is
a somewhat subtle point because of the momentum-dependent change of
variables (\ref{nonlocal}).}

Thus, the dynamics of the planar master field of $YM_{2+1}$ which
describes the vacuum in the planar, low momentum limit, in the
well-defined  momentum expansion of the reduced Hamiltonian, given in
terms of $\phi$ matrix variables, is determined by the following {\it
one} matrix model Hamiltonian
\begin{equation}
\frac{1}{2} \int Tr \left( \pi^2 +
m_r^2 \phi^2+...\right) \equiv
\frac{1}{2} \int Tr \left( -\frac{\delta^2}{\delta \phi^2} +
m_r^2 \phi^2+...\right)
\end{equation}
{\it The vacuum in this approximation is captured by the singlets,
invariant under the residual {\em unitary} transformation $\phi \to h
\phi h^{\dagger}$.} This is a huge reduction in the number of states in
the large $N$ limit, described by the density of $N \to \infty$
eigenvalues of the matrix $\phi$!

Therefore, due to the holomorphic invariance of the vacuum, the spectrum
of the planar limit of $YM_{2+1}$ is determined in the planar, low
momentum (large 't Hooft coupling) limit, by the spectrum of the singlet
states of this effective one matrix model.

Before proceeding further, let us note that in order to compute
correlation functions we would need to include momentum dependent terms
in the reduced Hamiltonian in terms of $\varphi$ matrix variables. One
obvious technical complication one has to deal with in order to get the
leading momentum expressions for various correlation functions is that
the momentum-dependent terms in the reduced Hamiltonian are not of the
usual kind considered in the literature of the one matrix model (i.e.
they are not traces or multi-trace terms involving a single matrix).
Consequently, the computation of correlation functions will be
significantly harder, although we believe the formalism presented here
is sufficient for a discussion of the spectrum. In this letter we
restrict our attention to the spectrum.

\subsection{The vacuum wave functional, one more time}

Now we demonstrate that the effective quadratic matrix model captures the
correct physics of the vacuum. First, we can easily find the correct
vacuum wave functional (which leads to the area law and a successful
empirical expression for the string tension) using the above matrix
field theory approach. Given the quadratic matrix field Hamiltonian
(\ref{eq:matrixfieldham}), we see that the ground state wave functional
is a Gaussian
\begin{equation}
\Phi = \exp\left(-\frac{1}{2}\int \phi\sqrt{m_r^2 - \nabla^2} \phi\right)
\end{equation}
To compare to the previous discussion, we should consider the effect of
transforming back to the non-trivial inner product of the original
collective field theory. This amounts to a simple shift of the exponent
of the wave functional
\begin{equation}
\Psi = \exp\left(-\frac{1}{2}\int \phi(x) \left(-m_r + \sqrt{m_r^2 - \nabla^2}\right) \phi(y)\right)
\end{equation}
Conversion to the $J$ variables to linear order in $\phi$ fields gives a
wave functional which perfectly matches the expression for the
wave functional (\ref{eq:KNwavefn}), which, as we have seen above, leads
to successful predictions of the area law and the string tension.

Thus we explicitly see that in the large $N$, low momentum, that is,
large 't Hooft coupling limit, the effective quadratic matrix model
leads to the correct physics of the Yang-Mills vacuum, as described by a
vacuum wave functional consistent with the area law. This result should
not come as too much of a surprise, since it is essentially the same
calculation that we considered in Section 2.1. However, it is a useful
check that the reduction has not thrown away anything important about
the vacuum.

We are now ready to show that the same effective matrix model describes
the correct spectrum of excitations about the vacuum of the large $N$
$2+1$-dimensional Yang-Mills theory.

\section{The planar spectrum of $YM_{2+1}$}

In this section we determine the planar spectrum of $YM_{2+1}$ from the
spectrum of the singlet states of a one-matrix model. The one matrix
model is a well-studied system \cite{onematrix}. It can be understood
from many points of view of the known large $N$ technology
\cite{collective}\cite{onematrix}\cite{master}. For example, it is well
known that the planar limit of a quadratic matrix model in the singlet
sector is completely captured by the semicircular Wigner-Dyson
distribution of the density of eigenvalues ($\lambda$) of the matrix
$\phi$
\begin{equation}
\label{wigner}
\rho_{WD}(\lambda)=\frac{1}{\pi}\sqrt{2 \alpha - m_r^2 \lambda^2}
\end{equation}
where $\lambda \in [-\Lambda, \Lambda]$, $\Lambda$ being the point where
the square root vanishes, that is $\Lambda = \frac{2\alpha}{m_r^2}$.
($\alpha$ is just a Lagrange multiplier associated with the
normalization condition satisfied by the eigenvalue density. In the
equivalent fermionic formulation, it is the Fermi energy.) Furthermore,
the master field of a Gaussian matrix model \cite{master} can be
described in terms of noncommutative probability theory \cite{ncpt} (for
which the semicircle distributions play the role completely analogous to
the Gaussian distributions of the commutative probability theory) as an
operator $ \phi  = a + \frac{\alpha}{2} a^{\dagger} $ acting on a Fock
space build out of $a$ and $a^{\dagger}$ operators and the vacuum
$|0\rangle$, $a |0\rangle =0$.  $a$ and $a^{\dagger}$ satisfy the Cuntz
algebra $ a a^{\dagger}  = 1. $

The spectrum on singlet excitations can be easily determined by
perturbing around the ground state value, determined by the semicircle
law, assuming $e^{i\omega  t}$ time dependence and reading off the
normal modes from the resulting wave equation. (For example, this has
been done in detail in \cite{shapiro} for the case of the one-matrix
model.) Because these calculations are crucial for our main claim, we
review them from a couple of different points of view, mainly following
the work of \cite{shapiro}.

\subsection{Spectrum of singlets from the collective field theory}

As stated above, the density of eigenvalues $\rho(\lambda)$ is the
correct variable that describes the dynamics of the large $N$ one matrix
model. One introduces first $\rho_k \equiv Tr \exp(-ik\phi)$, and then
defines the collective field $\rho(\lambda)$ as the Fourier transform of
$\rho_k$. Here $\rho(\lambda)$ is positive definite and normalized as
$\int \rho(\lambda) d\lambda =N$.

Suppose that the one matrix model is described by the general
Hamiltonian $Tr(\frac{1}{2} \pi^2 + V(\phi))$. The collective field
Hamiltonian for $\rho(\lambda)$ and its conjugate momentum $\Pi
(\lambda)$ reads as follows
\begin{equation}
\frac{1}{2} \int d\lambda [\rho(\lambda) (\partial_{\lambda} \Pi(\lambda))^2
+ \frac{1}{6} \pi^2 \rho(\lambda)^3 + V(\lambda) \rho(\lambda)-\alpha\rho(\lambda)]+\alpha N
\end{equation} 
The classical equations of motion for the collective field theory are
\begin{equation}
\label{eom}
\dot{\rho}(\lambda) = -\partial_{\lambda}
[\rho(\lambda)\partial_{\lambda} \Pi(\lambda)], \quad
\dot{\Pi}(\lambda) = \frac{1}{2}(\partial_{\lambda} \Pi(\lambda))^2 +
\frac{1}{2} \pi^2 \rho(\lambda)^2 + V(\lambda)-\alpha
\end{equation}
The solution of these equations that satisfies the
positive definiteness of the eigenvalue density is such that
\begin{equation}
\partial_{\lambda} \Pi(\lambda)=0,\ \ \ \ \
\rho(\lambda)=\rho_{WD}(\lambda).
\end{equation}

The spectrum can be now determined by small perturbations around this
classical ground-state solution. In particular, introduce following
\cite{shapiro}, $\xi(\lambda) = \partial_{\lambda} \Pi(\lambda)$ and
$\rho(\lambda)^2 = \rho_0(\lambda)^2 +\eta(\lambda)$, and expand the
Hamilton equations (\ref{eom}) to first order in $\xi$ and
$\eta$. Assuming the periodic time dependence $\exp(-i \omega t)$ for
both perturbations, we may eliminate $\xi$ to get the following wave
equation for $\eta$
\begin{equation}
\label{wave}
\omega^2 \eta + \frac{\partial^2 \eta}{\partial q^2}=0
\end{equation}
where $q(\lambda) = \int_0^{\lambda} [2(\alpha - V(x))]^{-\frac{1}{2}}
dx$.

The boundary conditions are determined from the constraint on the
density of eigenvalues $\int \rho(\lambda) d \lambda = N$ and the
vanishing of $\rho_0$ for $\lambda = \pm \Lambda$, from which one infers
that $\frac{\partial \eta}{\partial q} =0$ for $q=q(\Lambda)$. This
implies that the wave equation (\ref{wave}) describes a finite length
string with free ends (for example, $\eta \sim \cos\left(n\pi
\frac{q}{q(\Lambda)}\right))$ from which we infer the spectrum
\begin{equation}
\label{spectrum}
\omega = n \omega_c
\end{equation}
where the fundamental frequency is given by
\begin{equation}
\omega_c = \frac{\pi}{2} \left(\int_0^{\Lambda}d\lambda\
[2(\alpha - V(\lambda))]^{-\frac{1}{2}} \right)^{-1}.
\end{equation}
In the case of a quadratic potential $V(x) = \frac{1}{2} x^2$, this can
be integrated precisely into a gap-like equation
\begin{equation}
\omega_c = \frac{\pi}{2} \left(\int_0^{\Lambda}d\lambda\
[2\alpha - m_r^2 \lambda^2]^{-\frac{1}{2}}\right)^{-1} = m_r
\end{equation}

Of course, the same result follows by directly perturbing the collective
field Hamiltonian and determining the normal modes of the quadratic part
\cite{shapiro}.

As is well know, the above result (\ref{spectrum}) can be readily understood from a
different, and somewhat more intuitive, point of view. As is well known \cite{onematrix}, \cite{joep} the singlet
sector of the one matrix model can be understood in terms of an
effective system describing N free spinless fermions in a quadratic
potential. The relevant Hamiltonian is
\begin{equation}
\frac{1}{2} p^2 + V(\lambda)
\end{equation}
The ground state in this picture corresponds to a filled Fermi sea, with
the corresponding Fermi energy $\alpha$ so that
\begin{equation}
N = \frac{1}{\pi} \int d\lambda (2(\alpha - V(\lambda))
\end{equation}
The density of states in the semiclassical limit is given by
\begin{equation}
\frac{1}{2\pi}\int d\lambda d p \delta(\alpha - (\frac{1}{2} p^2 + V(\lambda)))=
\frac{1}{2\pi}\int d\lambda (2(\alpha - V(\lambda))^{-\frac{1}{2}}
\end{equation}
For a quadratic potential $V(\lambda) = \frac{1}{2} \lambda^2$
the fundamental frequency $\omega_c =m$ is given as an inverse of
the density of states at the Fermi level. Of course, this result agrees with the
corresponding collective field theory result. 
Notice, that $\omega_c$ is nothing but the classical frequency
of the corresponding classical trajectory describing the motion of free
fermions in the effective potential $V(\lambda)$.
Finally, the level degeneracy is also easy to discuss in the fermionic picture.

The crucial point here is that the singlet spectrum of a one matrix
model has a gap, determined by the mass parameter $m$, is equidistant
and consists of free non-interacting excitations. Going back to the
planar limit of $YM_{2+1}$, we see that for the large value of the 't
Hooft coupling, the master field of the planar $YM_{2+1}$, a constant
$\infty \times \infty$ matrix left invariant under the residual
holomorphic transformations, is equivalent to the eigenvalue
distribution of the {\it singlet} sector of a quadratic one matrix
model. Therefore the spectrum has a gap determined by the mass $m$ and
is equidistant. {\it The mass gap emerges because of the finiteness of
the cut of the semi-circle distribution and the nature of the boundary
conditions at the end points of the cut.} This offers an interesting
view on confinement.

The Karabali-Nair formalism is completely consistent with Lorentz invariance
as shown in \cite{knair}.
The Lorentz invariant form of the planar $YM_{2+1}$ spectrum is then
given by the following dispersion relation describing an equidistant
spectrum with an explicit mass gap 
\begin{equation} 
\omega(\vec{k}) = \sqrt{{\vec{k}}^2 + m_n^2} 
\end{equation} where $n=1,2,...$, $m_n = n
m_r$, where $m_r$ is the renormalized gauge invariant mass and the bare
mass $m^2$ is determined by the planar 't Hooft coupling $g_{YM}^2 N$
via $m= \frac{g_{YM}^2 N}{2 \pi}$. First of all, this formula elucidates
the meaning of the parameter $m$ in the planar limit of $YM_{2+1}$ in
the large 't Hooft coupling limit. Note that as the 't Hooft coupling is
taken to zero we recover the dispersion relations for massless
particles. This is in accordance with the recovery of a Coulomb
potential at short distances, as shown in \cite{knair}. Finally, as we
have seen, the matrix model knows about the vacuum wave functional which
in the high momentum limit does describe a theory with a Coulombic
potential and in the low momentum limit leads to the area law for the
expectation value of the Wilson loop and an explicit formula for the
string tension which matches the numerical data.

\subsection{Matrix model reduction and the planar spectrum}

Obviously, the structure of higher order momentum dependent terms in the
effective gauge invariant Hamiltonian is in general 
non-local and very involved. The question
that we would like to briefly consider is how the higher order terms
influence the leading quadratic result for the mass gap and the
spectrum. Even though the
higher order corrections are explicitly non-local, 
we will argue that the existence of a mass gap
in the large N, large 'tHooft coupling limit,
implies a self-consistent expression for the
mass term even upon the inclusion of the
higher order terms in $\varphi$.

As we have seen, the planar vacuum at leading
order in the 'tHooft coupling is governed 
by the quadratic part of the
collective field Hamiltonian, and the
higher order contributions to the wave functionals
can be re-organized into an
expansion in inverse powers of $m$.
Motivated by this observation, suppose we 
cut-off the momentum integrals in all non-local terms
in the collective field Hamiltonian
by the mass gap $M$ and then expand in momentum
and finally perform the large
N reduction. This procedure is difficult to
implement technically, yet nevertheless, 
the general form of the effective reduced Hamiltonian
involving $\phi$ matrices, would look as follows
\begin{equation}
\frac{1}{2} \int Tr \left( -\frac{\delta^2}{\delta \phi^2} +
V(\phi, m_r)+...\right)
\end{equation}
The coefficients in $V(\phi)$ would only involve powers of the {\it
renormalized} Karabali-Nair mass parameter $m$, denoted by
$m_r$.

Now, by insisting on the holomorphic invariance of the vacuum, as in the
quadratic case, we see that only the singlet sector of this general one
matrix model, corresponds to the vacuum of $YM_{2+1}$. The spectrum of
the singlets has been discussed in the previous section. The fundamental
frequency, corresponding to the gap of $YM_{2+1}$ is given by
\begin{equation}
\omega_c = \frac{\pi}{2} (\int_0^{\Lambda(m_r)}
[2(\alpha - V(\lambda, m_r))]^{-\frac{1}{2}} d\lambda)^{-1}
\end{equation}
By choosing $\omega_c$ to correspond to the physical mass gap $M$,
and by evaluating the renormalized mass at the scale
determined by $M$, we get a self-consistent {\it gap equation}
\begin{equation}
M = \frac{\pi}{2} (\int_0^{\Lambda(m_r(M))}
[2(\alpha - V(\lambda, m_r(M)))]^{-\frac{1}{2}} d\lambda)^{-1}
\end{equation}
Because of the positive definite nature of the potential,
a solution to this equation should exist.
The lowest numerical value, in the units of the 't Hooft coupling
should correspond to the physical gap of $YM_{2+1}$.

One important point here is that the lowest order, quadratic
contribution is not misleading when we try to capture the correct long
distance and short distance physics in the planar limit. The ultimate
reason for this is that the gauge invariant $H$ variables are local, and
that the inner product on the space of states is computable in terms of
these variables. The lowest, quadratic order for the master field
(described by the dynamics of an effective quadratic matrix model)
obviously captures an important piece of the correct ground state (i.e.
the master field). The effective matrix model leads to the ground state
that is also consistent with the form of the vacuum wave functional
obtained in \cite{knair}, predicting the string tension which turns out
to be in excellent agreement with the available lattice data
\cite{teper}. In our case, the numerical value of the mass gap as given
by the spectrum of singlets of the effective matrix model has to be
fitted to the lattice data.

\subsection{Towards the QCD String}

Finally we add a couple of comments regarding the relevance of
our discussion for a possible string theory of the $2+1$ dimensional
Yang Mills theory.

Perhaps the two obvious questions regarding our results in
view of the usual intuition connected to the QCD string are \footnote{We thank
Ofer Aharony, David Berenstein and Shiraz Minwalla for 
discussions regarding these issues.}:
1) how would one demonstrate, within our framework, the expected Hagedorn behavior of the
density of states \cite{hagedorn}
and also 2) how would one establish the existence of Regge trajectories
expected from a QCD string?

To approach answering these questions recall that the Karabali-Nair variables
capture all the degrees of freedom of the $2+1$ dimensional
Yang Mills theory. Also, the Karabali-Nair Hamiltonian is the
{\it exact} collective field Hamiltonian.
Thus, in the confined phase of the theory one expects the
usual $O(1)$ scaling of the free energy and in the deconfined
phase the $O(N^2)$ behavior.
Indeed, the effective matrix model we have used to demonstrate
the existence of the mass gap, corresponds to the counting of
degrees of freedom expected from a confining phase.
On the other hand, the original Karabali-Nair collective field
Hamiltonian does describe $O(N^2)$ perturbative degrees of freedom.
The presence of an explicit mass gap, indicates that the 
self-consistent
effective scalar matrix field theory derived from the Karabali-Nair
collective field description is cut off at the scale 
determined by the gap $M$, which would be in accordance with
the expected confinement/deconfinement transition.

In order to really establish the stringy Hagedorn behavior (and also
demonstrate the existence of Regge trajectories) in the planar 
$YM_{2+1}$ one needs to attach the Lorentz indices to the
oscillators of the singlet sector of the effective matrix model.
The effective matrix model already describes the correct vacuum
in the planar limit and the oscillators acting in the corresponding
Fock space should obey the algebra of the usual oscillators
of the free string theory, because of the large N factorization.
This we think is the key to deriving a QCD string field theory for
the $2+1$ dimensional
Yang Mills theory and the associated stringy (such as Hagedorn and Regge)
features. Work on this important issue is in progress.

\section{Concluding remarks}

To conclude, the spectrum of physical excitations in the planar limit of
$YM_{2+1}$ can be deduced using the remarkable local holomorphic variables of Karabali and Nair taken in conjuction with some well
known results from large $N$ master field technology. The analytic
understanding of the spectrum is possible due to a reduction of the
$YM_{2+1}$ Hamiltonian for the large 't Hooft coupling (low momentum
limit) to the {\it singlet} sector of an effective one matrix model. The
huge reduction of the degrees of freedom to the singlet sector of the
one matrix model is a consequence of the holomorphic invariance of the
$YM_{2+1}$ vacuum in the Karabali-Nair representation. Note also that
the matrix model captures the form of the vacuum wave functional which
leads to the area law and a successful empirical expression for the
string tension, reinforcing the self-consistency of our approach.

Obviously the approach presented in this letter is just a first step in
the direction of unraveling the full planar limit of $2+1$ dimensional
Yang-Mills theory. Given our result we believe that we are now well
motivated to study the full matrix collective field theory for the gauge
invariant holomorphic loop variables. We also believe that some of the
old as well as currently pursued ideas pertaining to the subject of the
QCD string \cite{polyakov} such as the question of integrability, the
analogy between gauge and chiral fields, as well as target space
understanding of the gauge theory/gravity duality should be re-examined
in the context of $2+1$ dimensional Yang-Mills theory. The gauge
invariant collective field Hamiltonian can be understood as a string
field theory of $YM_{2+1}$. It would be obviously very interesting to
understand the world-sheet structure underlying this target space
description in order to get even closer to the (perhaps not so) elusive QCD
string. Last, but not least, we believe that a new way is open for a
rational analytic approach to the large $N$ $3+1$ dimensional Yang-Mills
theory.

\section*{Acknowledgments}

First we would like to thank V.P. Nair for patient explanations of his
work with Dimitra Karabali. We also want to thank Joe Polchinski for
important discussions and for pushing us to write up a preliminary
version of our ideas as well as Vishnu Jejjala for his participation in
the early stages of this project. Finally, very special thanks to David
Gross for his interest in this work and a very careful reading of the
preliminary draft of this paper. {\small RGL} was supported in part by
the U.S. Department of Energy under contract DE-FG02-91ER40709.{\small
DM} was supported in part by NSF grant PHY-9907949 at KITP, Santa
Barbara and by the U.S. Department of Energy under contract
DE-FG05-92ER40677. {\small RGL} and {\small DM} acknowledge the
hospitality and support of the Aspen Center for Physics
and the CERN Theory Division. Finally,
{\small DM} would like to acknowledge the stimulating atmospheres of the Kavli Institute for Theoretical Physics, Berkeley Center for Theoretical Physics, and Perimeter Institute for Theoretical Physics,
where various portions of this work were
done.

\section*{Appendix}

Here we collect some notation and results. First, define $G_{(x,y)}$ and
$\bar G_{(x,y)}$
\begin{equation} \partial G_{(x,y)}=\delta^{(2)}(x-y), \ \ \ \ \ \bar\partial\bar G_{(x,y)}=\delta^{(2)}(x-y)\end{equation}
\begin{equation} G_{(x,y)}=\frac{1}{\pi}\frac{1}{\bar x-\bar y},\ \ \ \ \ \bar G_{(x,y)}=\frac{1}{\pi}\frac{1}{x-y}\end{equation}
Gauge covariant versions are
\begin{equation} D^{-1}_{(x,y)}=(\partial+A)^{-1}_{(x,y)}=M(x)M^{-1}(y)G_{(x,y)}\end{equation}
\begin{equation}\bar D^{-1}_{(x,y)}=(\bar\partial+\bar A)^{-1}_{(x,y)}=M^{-\dagger}(x)M^{\dagger}(y)\bar G_{(x,y)}\end{equation}

We thus find
\begin{equation} \delta M(x)=-\int_y D^{-1}_{(x,y)}\delta A(y)M(y)=-M(x)\int_y G_{(x,y)}(M^{-1}\delta AM)(y) \end{equation}
\begin{equation} \delta M^\dagger(x)=\int_y M^\dagger(y)\delta \bar A(y)\bar D^{-1}_{(y,x)}=\int_y(M^{\dagger}\delta \bar A M^{-\dagger})(y) \bar G_{(y,x)}M^\dagger(x) \end{equation}

\begin{equation} \delta J=-\frac{c_A}{\pi}(M^\dagger\delta AM^{-\dagger})(x)+\frac{c_A}{\pi}\int_y(M^{\dagger}\delta \bar A M^{-\dagger})(y)\partial_x\bar G_{(y,x)}-\int_y\left[ J(x),(M^{\dagger}\delta \bar A M^{-\dagger})(y)\right]\bar G_{(y,x)}\end{equation}
This equation is used to derive ${\cal H}_{KN}[J]$ together with the
regulated expression $\Tr t^a \bar D^{-1}(x,x)=\frac{1}{\pi}\Tr
t^a(A-M^{-\dagger}\partial M^\dagger)$.

It also follows that
\begin{equation} \delta H(x)=-H(x) \int_y G_{(x,y)}(M^{-1}\delta AM)(y)+\int_y (M^{\dagger}\delta \bar A M^{-\dagger})(y)\bar G_{(y,x)}H(x)\end{equation}
But we also have
\begin{equation}\delta H=Ht^a\delta\varphi^b (e_{[\varphi]})_{ba}=t^a H \delta\varphi^b(e_{[\varphi]})_{ab}\end{equation}
and so
\begin{equation} \delta\varphi^a=2(e^{-1}_{[\varphi]})_{ba}\Tr t^b H^{-1}\delta H=2(e^{-1}_{[\varphi]})_{ab}\Tr t^b\delta HH^{-1}\end{equation}
Thus we conclude
\begin{equation}\frac{\delta\varphi^a(x)}{\delta A^b(y)}=2i(e^{-1}_{[\varphi]})_{ca}(x)G_{(x,y)}\Tr t^cM^{-1}(y)t^bM(y)\equiv -iM^{bc}(y)G_{(y,x)}(e^{-1}_{[\varphi]})_{ca}(x)\end{equation} 
\begin{equation}\frac{\delta\varphi^a(x)}{\delta \bar A^b(y)}=-2i(e^{-1}_{[\varphi]})_{ac}(x)\Tr t^cM^{\dagger}(y)t^bM^{-\dagger}(y)\bar G_{(y,x)}\equiv i(e^{-1}_{[\varphi]})_{ac}(x)\bar G_{(x,y)}{M^\dagger}^{cb}(y)\end{equation} 

\begin{equation} \frac{\delta^2 \varphi^a(x)}{\delta A^b(y)\delta \bar A^c(z)}= {M(y)}^{bd}G_{(y,x)}(e^{-1})_{da,g}(e^{-1})_{ge}\bar G_{(x,z)}{M^\dagger(z)}^{ec}\end{equation}
Introducing a gauge-invariant point-splitting procedure (insertion of a point-splitting Wilson line), we can use these expressions in the evaluation of the Hamiltonian
\begin{eqnarray} &&-\frac{g_{YM}^2}{2}\left[\int_{x,z} W^{cb}(z,x)\frac{\delta^2\varphi^a(y)}{\delta A^c(z)\delta\bar A^b(x)}\frac{\delta}{\delta\varphi^a(x)} +\int_{x,y,z} 
W^{dc}(z,x)\frac{\delta\varphi^b(w)}{\delta A^d(z)}\frac{\delta\varphi^a(y)}{\delta\bar A^c(x)}\frac{\delta^2}{\delta\varphi^b(w)\delta\varphi^a(y)}\right]\\
\simeq &&-m\left[ \int_x \varphi^a(x)\frac{\delta}{\delta\varphi^a(x)}-\int_{x,y} C(x,y)\frac{\delta^2}{\delta\varphi^a(x)\delta\varphi^a(y)}+\ldots\right]
\end{eqnarray}

As indicated in the text, another way to proceed is to work with ${\cal H}_{KN}[J]$ and convert to the $\varphi$ variables. We then need
\begin{eqnarray}
\frac{\delta}{\delta\varphi^c}&=&\frac{c_A}{i\pi}\left[ e_{ac}\partial-(e_{ab,c}-e_{ac,b})\partial\varphi^b\right]\frac{\delta}{\delta J^a}\\
&=&\frac{c_A}{i\pi} D'_{ca}\frac{\delta}{\delta J^a}
\end{eqnarray}
(note that
${D'}_{ab}\simeq\delta_{ab}\partial-\frac{i}{2}f_{abc}\varphi^c\partial+if_{abc}\partial\varphi^c+\ldots$),
so
\begin{equation}\frac{\delta}{\delta J^a(x)}=\frac{i\pi}{c_A}\int_y {{D'}^{-1}_{ac}}_{(x,y)}\frac{\delta}{\delta\varphi^c(y)}\end{equation}

\end{document}